\documentstyle[epsfig, emulateapj5]{aastex} 
\def\msun{{\rm\,M_\odot}} 

\def\zsun{{\rm\,Z_\odot}}
\newcommand{\etal}{et al.\ }

\def\h2{${\rm\,H_2}$}

\makeatletter 
\newenvironment{figurehere} {\def\@captype{figure}} {}

\begin{document}
\title{Gamma-Ray Bursts May Be Biased Tracers of Star Formation}

\author{Renyue Cen\altaffilmark{1} and Taotao Fang\altaffilmark{2}}

\begin{abstract}

Based on a simulation of galaxy formation in the standard 
cosmological model, we suggest that a consistent picture 
for Gamma-Ray Bursts and star formation may be found
that is in broad agreement with observations:
{\it GRBs preferentially form in low metallicity environments
and in galaxies substantially less luminous that $L_*$.}
We find that the computed formation rate
of stars with metallicity less than $0.1\zsun$
agrees remarkably well with the rate evolution 
of Gamma-Ray Bursts observed by Swift from $z=0$ to $z=4$,
whereas the evolution of total star formation rate  
is weaker by a factor of about $4$.
Given this finding, we caution 
that any inference of star formation rate based 
on observed GRB rate may require a more involved exercise than
a simple proportionality.
\end{abstract}


\altaffiltext{1}{Department of Astrophysical Sciences, Princeton
  University, Peyton Hall, Ivy Lane, Princeton, NJ 08544}

\altaffiltext{1}{Department of Astronomy, University of California,
Irvine, CA 92697; {\sl Chandra} Fellow}

\section{Introduction}

The intriguing observational linkage between long duration Gamma-Ray Bursts (GRBs)
 and core-collapse supernovae
(e.g., Stanek \etal  2003; Hjorth \etal 2003) suggests
that the progenitors of GRBs may be very massive stars.
This possible connection was predated by a proposed unified picture
(Cen 1998).
As such, it has been hoped that GRBs may be a good tracer of cosmic
star formation (Wijers \etal 1998;
Totani 1999;
Lamb \& Reichart 2000;
Blain \& Natarajan 2000;
Porciani \& Madau 2001; Daigne \etal 2006; Coward 2006;
Le \& Dermer 2007; Li 2007).
However, recent observations indicate that typical GRBs may prefer
relatively low metallicity environments ($\sim 0.1\zsun$)
(Fynbo \etal 2003;
Le Floch \etal 2003;
Christensen \etal 2004;
Fruchter \etal 2006;
Stanek \etal 2006)
and host galaxies significantly less luminous than $L_*$ 
(Fruchter \etal 1999,2006), 
although there is
evidence that the actual spread in metallicity may be
wide (Berger \etal 2006; Prochaska 2006; Wolf \& Podsiadlowski 2007).

The aim of this {\it Letter} is to first address
the issue of consistency of GRB environment with respect to
metallicity and galaxy luminosity,
i.e., the galaxy luminosity-metallicity relation,
in the context of detailed simulation of galaxy formation in
the standard cosmological model.
Then, we make predictions on the evolution of GRB rate with
redshift and highlight a possible dramatic difference between overall star formation
history and GRB rate history, if GRBs are not an unbiased tracer of star formation.
In particular, if GRBs are predominantly produced by stars with
metallicity $\le 0.1\zsun$, the GRB rate is expected in our model to
rise obstinately from $z=0$ to $z\sim 5$ by a factor of
$\sim 100$, when it flattens out towards higher redshift,
whereas the overall star formation rate rises rapidly only from $z=0$ to $z\sim 3$
and is roughly flat from $z\ge 3$ until $z\sim 7$.
The evolution of GRB rate with redshift is thus expected to be 
stronger than that of star formation.

\section{Evolution of GRB and Star Formation Rates}

Observations seem to indicate that GRB galaxy hosts are preferentially
dwarf galaxies of about $0.1L_*$ 
at low redshift (Fruchter \etal 1999,2006). 
This would be surprising, if GRB rate 
is directly proportional to total star formation rate,
because the latter is known to peak in significantly larger galaxies
(see, for example, Figure 1 below).
The implication is that GRB rate is not exactly proportional to 
the overall star formation rate.
On the other hand, analysis of the metallicity of GRB progenitors 
suggests that GRB progenitors 
tend to have relatively lower metallicity of $\sim 0.1\zsun$ than
that of typical forming stars at low redshift
(Fynbo \etal 2003;
Le Floch \etal 2003;
Christensen \etal 2004;
Fruchter \etal 2006;
Stanek \etal 2006).
We would like to ask the following question: 
are these two observational facts consistent 
in the galaxy formation model 
in the standard cosmological model?

Figure 1 shows the distribution of SDSS U-band light
as a function of the stellar mass of galaxies,
with the galaxies being divided into two subgroups
according to the mean metallicity of galaxies.
The results are based on a cold dark matter cosmological simulation
with galaxy formation 
in a cold dark matter universe
(Nagamine \etal 2006;
Cen \& Ostriker 2006; Cen \& Fang 2006)
with the following essential parameters:
$\Omega_M = 0.31$,
$\Omega_b = 0.048$, 
$\Omega_\Lambda=0.69$, $\sigma_8 = 0.89$,
$H_0 = 100 h$km s$^{-1}$ Mpc$^{-1}=69$ km s$^{-1}$ Mpc$^{-1}$ 
and $n_s = 0.97$.
The simulation box size is $85 h^{-1}$Mpc 
comoving with a number of cells of 
$1024^3$, giving a cell size of 83~h$^{-1}$ kpc comoving
and dark matter particle mass equal to 
$3.9 \times 10^8 h^{-1}\msun$.
Given a lower bound of the temperature for almost all the gas in the simulation of 
$T\sim 10^4$ K, the Jeans mass is 
$\sim 10^{10}\msun$ for mean-density gas, 
which is comfortably larger than our mass resolution.
Galaxies are produced using a grouping scheme HOP (Eisenstein \& Hut1998)
(see Nagamine \etal 2001 for details),
which gives a catalog of galaxies each with stellar mass,
mean stellar metallicity, luminosities in all SDSS bands, etc.

We see a general trend in Figure 1
that larger galaxies tend to have higher metallicity,
albeit a significant dispersion exists (not shown here).
Specifically, it is seen that, if the stellar metallicity 
is, say, typically lower than $0.1\zsun$ for GRBs progenitors,
galaxies of stellar mass $10^{10}\msun$ is expected to make the dominant contribution,
whereas the overall star formation rate is peaked in galaxies
of stellar mass $10^{11}-10^{12}\msun$.
Apparently, this is in broad agreement with observations of GRB
being concentrated in dwarf galaxies of metallicity $<0.1\zsun$.
This is reassuring in that our simulation that has been shown to produce
consistent results compared to the real universe in many other aspects
appears to be in broad agreement with observations with regard to 
the general luminosity-metallicity trend
(e.g., Kobulnicky \& Kewley 2004)
and with the observed preference of GRB hosts
of typical metallicity $\sim 0.1\zsun$ and typical luminosity
$\le 0.1L_*$.

\begin{figurehere}
\begin{center}
\resizebox{3.5in}{!}{\includegraphics[angle=90]{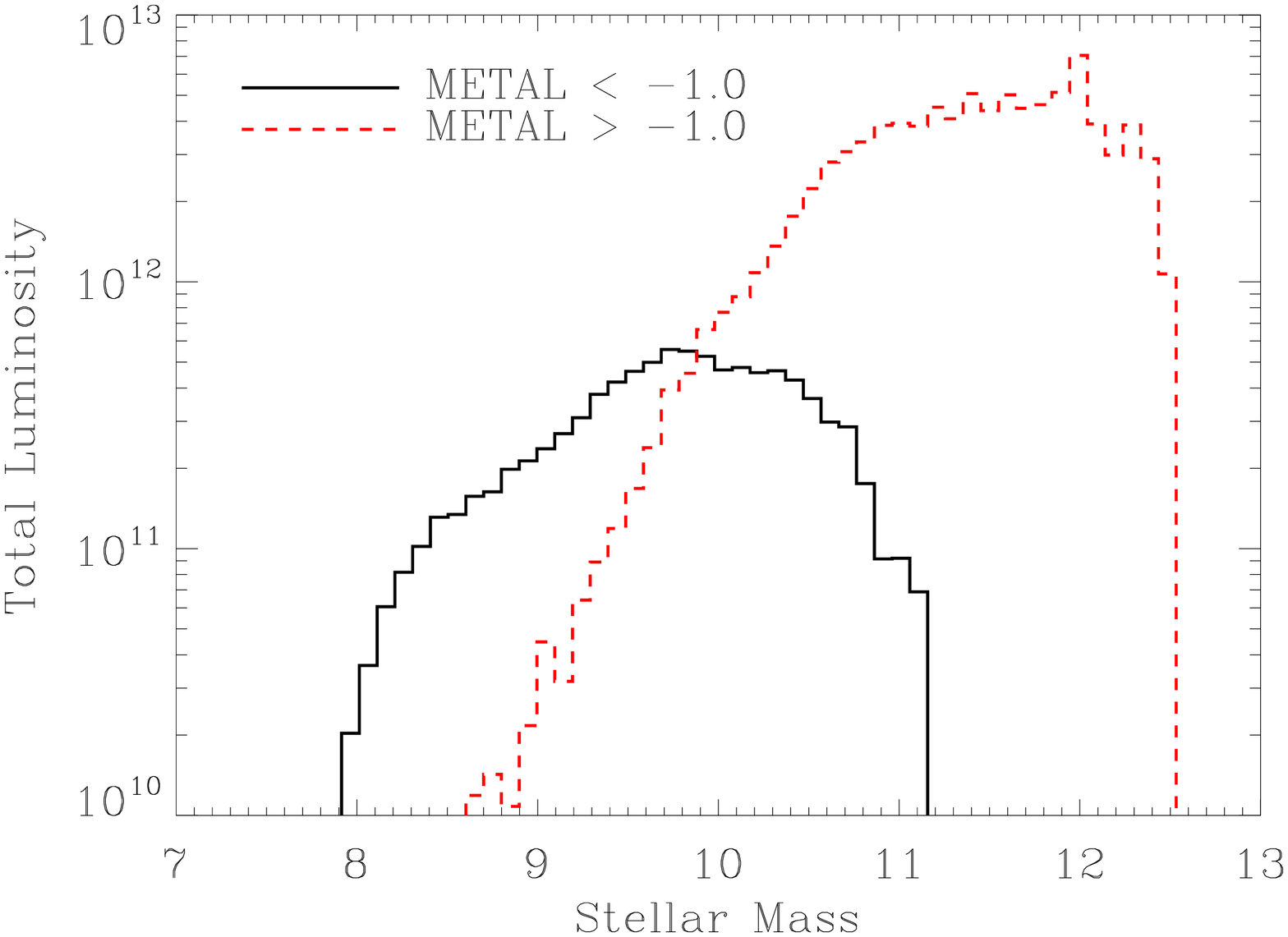}}
\end{center}
\caption{
shows the distribution of U band light
as a function of the stellar mass of galaxies,
for two groups of galaxies,
one with metallicity greater than 10\% of solar metallicity (dashed histogram)
and the other with metallicity lower than 10\% of solar metallicity (solid histogram).
}
\label{fig:sfr}
\vskip7pt
\end{figurehere}

Having verified that our simulation
is able to reproduced the general luminosity-metallicity relation for galaxies,
we will take a step further to infer the rate evolution for GRBs,
given the computed star formation rate and stellar metallicity as a function of redshift.
Since our simulated galaxies are composed of 
thousands to millions of ``stellar" particles,
which typically resemble globular clusters,
we will use the ``resolved" metallicity of individual ``stellar" particles.
Usually, there is a wide dispersion in stellar metallicity among
the ``stellar" particles within an individual galaxy,
reflecting complicated star formation history of each galaxy.

Figure 2 shows histories of three rates: total star formation (dotted curve),
star formation with metallicity less than $0.3\zsun$ (dashed curve)
and less than $0.1\zsun$ (solid curve), respectively.
We see that, if GRBs are preferentially hosted by stellar progenitors
with metallicity lower than $0.1\zsun$, 
one should expect to see their rate 
to rise approximately exponentially as a function of redshift
from $z=0$ until $z\sim 5$.
This is contrasted with a flattening of the overall star formation rate
at a much lower redshift $z\sim 2$ accompanied by only a modest rise (a factor of $2$)
from $z\sim 2$ to $z\sim 5$.

\begin{figurehere}
\begin{center}
\resizebox{3.5in}{!}{\includegraphics[angle=90]{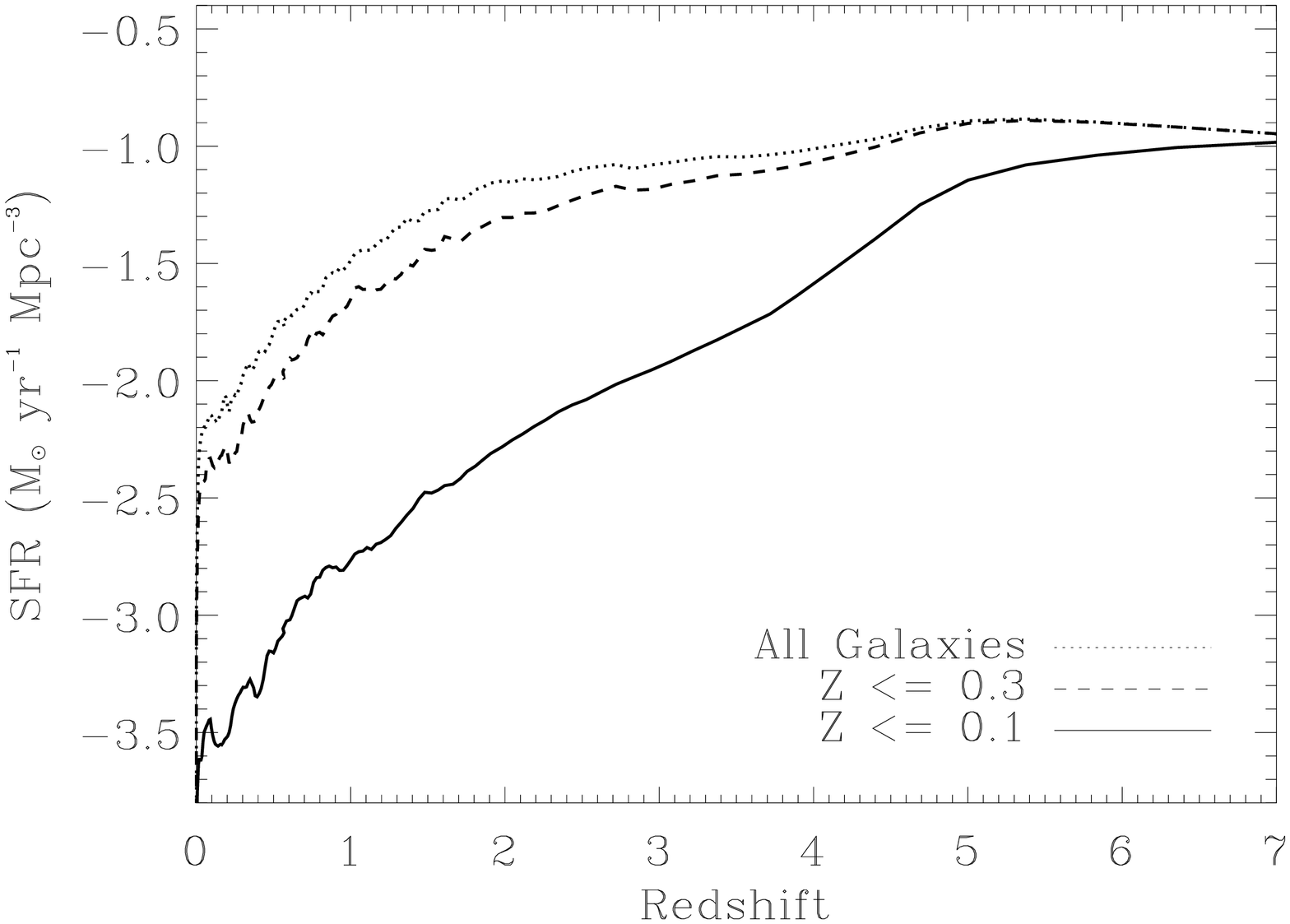}}
\end{center}
\caption{
shows histories of three rates: total star formation (dotted curve),
star formation with metallicity less than $0.3\zsun$ (dashed curve)
and less than $0.1\zsun$ (solid curve), respectively.
}
\label{fig:sfr}
\vskip7pt
\end{figurehere}

We note that, while the overall star formation seen in Figure 2
is higher by a factor of 
$\sim 30$ at $z=4$ than at $z=0$,
the ratio of GRB rate at $z=4$ to that at $z=0$ is $\sim 120$,
 if GRB progenitors 
predominantly have metallicity less than $0.1\zsun$,
roughly a factor of four larger.
This result is, curiously, in remarkable agreement
with recent observations of Kistler \etal (2007) where
they suggest that the GRBs observed by Swift
has an enhanced evolution by a factor of $\sim 4$ from $z=0$ to $z=4$
compared to the overall star formation rate.

\begin{figurehere}
\begin{center}
\resizebox{3.5in}{!}{\includegraphics[angle=90]{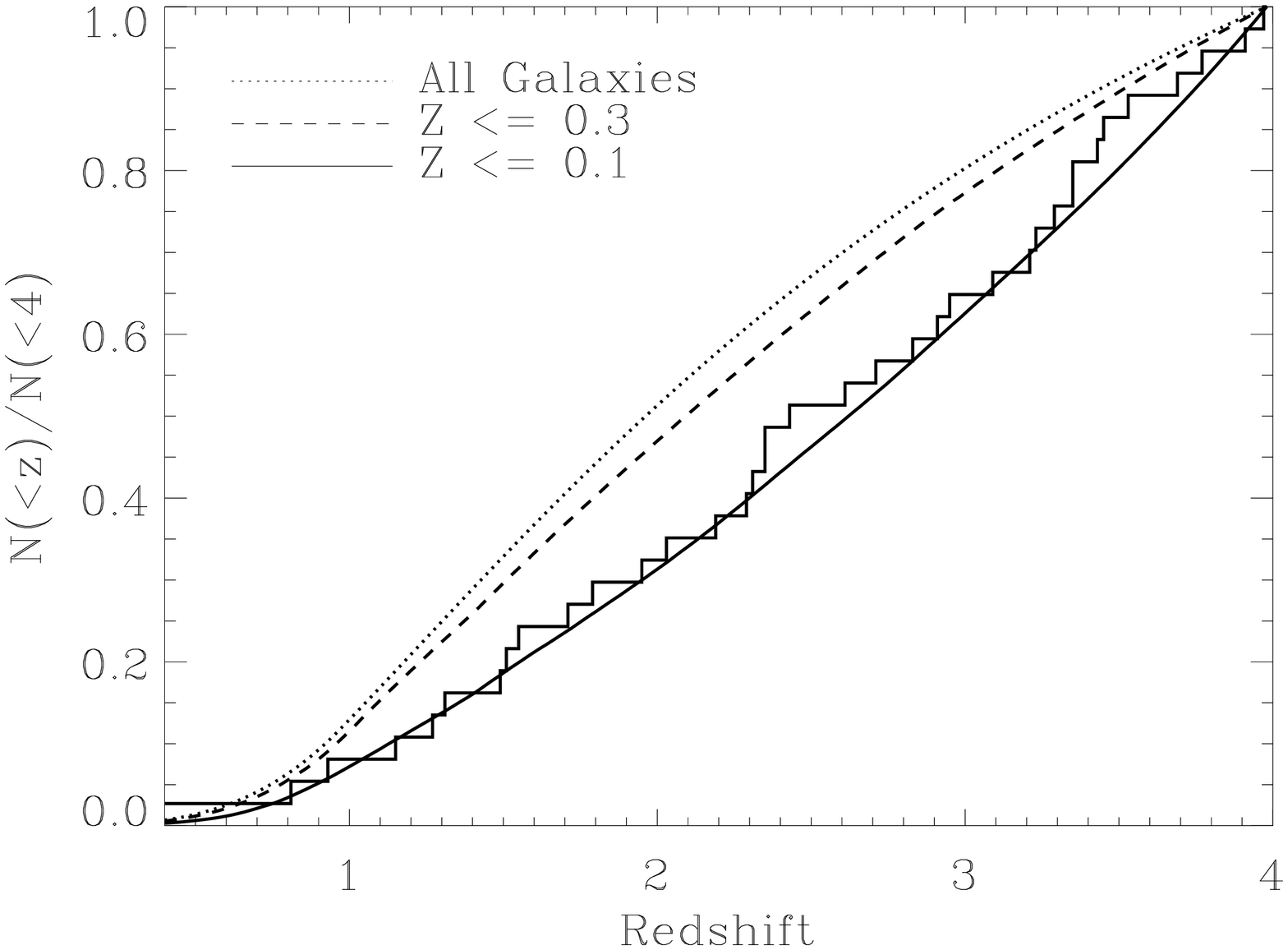}}
\end{center}
\caption{
shows the cumulative distribution of the observed (steps) and predicted (lines) GRB rates, normalized to the total number of bursts between z=0 and z=4. See Fig. 2 for line styles. Data shows the distribution of the 36 $\sl Swift$ GRBs taken from Butler et al.~(2007). The star formation rate of all-galaxy sample was normalized to the
analytic values of Hopkins \& Beacom (2006), and the star formation rates for the other two samples was normalized by the same factor.
}
\label{fig:sfr}
\vskip7pt
\end{figurehere}

In the spirit of making as a direct comparison as possible
with observations,
we shows the cumulative distribution of the observed (steps) and predicted (curves) GRB rates, normalized to the total number of bursts between $z=0$ and $z=4$ in Figure 3, following Kistler \etal (2007). 
Here, in order to make a direct comparison with Kistler \etal (2007)
and to remove uncertainties in the overall star formation
rate in our simulation,
we normalize 
the star formation rate of the all-galaxy sample in our simulation
to the analytic values of Hopkins \& Beacom (2006), 
as did  Kistler \etal (2007).
Then, the star formation rates for the other two (lower metallicity)
samples are adjusted multiplicatively by the same factor. 
It is seen that,
if one simply assumes that
the total GRB rate is proportional to 
the star formation rate of metallicity less than $0.1\zsun$,
the observed evolution of GRB rate from $z=0$ to $z=4$ is reproduced
to a high degree,
clearly visible in the good agreement between the solid curve
and the step curve in Figure 3.

\section{Discussion}

Taken all the observational facts together along with
our theoretical results, 
a broadly consistent picture appears to emerge:
{\it GRBs preferentially form in low metallicity environments
and (because of the luminosity-metallicity relation)
in galaxies substantially less luminous that $L_*$,
and GRB rate evolves more strongly than the overall star formation rate.} 
Evidently, 
the observed evolution of GRB rate from $z=0$ to $z=4$ 
can be explained, {\it if 
GRBs are primarily produced by massive
stars with metallicity less than $0.1\zsun$}.
Including higher metallicity stars 
would produce GRB rate evolution from
$z=0$ to $z=4$ that is inconsistent with observations.
Nevertheless, this overall picture would also be consistent
with a theoretical preference or possibly requirement
 of low metallicity for GRB progenitors
in the context of ``collapsar" models (MacFayden \& Woosley 1999; 
Woosley \& Heger 2006).

What is implicitly assumed is that the stellar 
initial mass function (IMF) 
has remained the same over the redshift range considered.
In other words, whatever metallicity dependence
of GRB rate may have,
this dependence is assumed not to evolve with redshift.
One should note that it is not fully known observationally or 
understood theoretically how the IMF evolves with time.
Therefore, additional possible evolutionary effect of IMF
would add another layer of complexity to this issue.
It is often thought 
that lower metallicity environment might 
favor formation of more massive stars.
If GRB progenitors are massive stars,
this would then translate to the expectation
that additional effect due to an evolving IMF
may further steepen the evolution of the GRB rate with redshift. 
This, however, is not required or borne out
in our analysis.
Our results thus imply that
the evolution of IMF from $z=4$ to $z=0$, if any,
appears to be weak.

If we place this result in a larger context of
star formation over the entire cosmic history,
one might come to the conclusion that,
while there may be a 
dramatic transition of IMF from
Population III metal-free stars 
(Nakamura\& Umemura, M. 2002;
Abel, Bryan \& Norman 2002;
Bromm, Coppi, \& Larson 2002)
to Population II stars at some high redshift 
(e.g., Cen 2003; Trac \& Cen 2007),
further evolution of IMF at lower redshift may be modest,
in the sense that the mass fraction of high mass stars
that are presumably GRB progenitors of the total
stellar mass remains relatively constant.

\section{Conclusions}

We utilize a simulation of galaxy formation in the standard 
cosmological model that has been shown to produce results 
consistent with extant observations of galaxy formation (e.g., Nagamine \etal 2006)
to shed light on the relation between GRB rate and star formation rate.
We find that a consistent picture 
for Gamma-Ray Bursts and star formation 
that is in broad agreement with observations
would emerge, 
{\it if GRBs preferentially form in low metallicity environments
and in galaxies substantially less luminous that $L_*$.}
Because of the increase of metallicity with cosmic time,
GRB rate consequently evolves more strongly 
with redshift than the overall star formation rate.
We find that 
the observed evolution of GRB rate from $z=0$ to $z=4$ 
can be explained, {\it if 
GRBs are primarily produced by massive
stars with metallicity less than $0.1\zsun$},
whereas an inclusion of stars with metallicity as high as $0.3\zsun$ 
yields GRB rate evolution from
$z=0$ to $z=4$ inconsistent with observations.

Therefore, we reach the conclusion
that GRBs may not be a good tracer of cosmic star formation,
especially over a long timeline.
As a result, a simple inference of star formation rate
or its derived quantities such as the ionizing photon production rate at high redshifts,
based on the observed GRB rate,
should be done with caution and may require careful calibrations.

\smallskip
\acknowledgements
{We thank Ken Nagamine for providing simulated galaxy catalogs.
We gratefully acknowledge financial support by
grants AST-0407176 and NNG06GI09G.}


\begin{thebibliography}{99}

\bibitem[Abel \etal (2002)]{a02}Abel, T., Bryan, G.L., \& Norman, M.L., 2002, Science, 295, 93


\bibitem[Berger \etal (2006)]{b06} Berger, E. \etal 2006, ApJ, 642, 979

\bibitem[Blain \& Natarajan (2000)]{bn00} Blain, A.W., \& Natarajan, P. 2000, MNRAS, 312, L35

\bibitem[Bromm \etal (2000)]{bn00}Bromm, V., Coppi, P.S., \& Larson, R.B. 2002, ApJ, 564, 23

\bibitem[Butler \& Natarajan (2007)]{bn00} Butler, N.R., Kocevski, D., Bloom, J.S. \& Curtis, J.L., arXiv:0706.1275, ApJ in press

\bibitem[Cen (1998)]{c98} Cen, R., 1998, ApJ, 507, L131

\bibitem[Cen (2003)]{c03} Cen, R., 2003, ApJ, 591, 12

\bibitem[Cen \& Fang (2006)]{cf06} Cen, R., \& Fang, T. 2006, ApJ, 650, 573

\bibitem[Cen \& Ostriker (2006)]{co06} Cen, R., \& Ostriker, J.P. 2006, ApJ, 650, 560

\bibitem[Christensen \etal (2004)]{c04} Christensen, \etal 2004;

\bibitem[Coward(2007)]{2007NewAR..51..539C} Coward, D.\ 2007, New Astronomy Review, 51, 539 

\bibitem[Daigne et al.(2006)]{2006MNRAS.372.1034D} Daigne, F., Rossi, E.~M., \& Mochkovitch, R.\ 2006, \mnras, 372, 1034 

\bibitem[Eisenstein and Hut(1998)]{Eisenstein98} Eisenstein, D. J.~and~Hut, P.  1998, \apj, 498, 137

\bibitem[Fruchter \etal (1999)]{f99} Fruchter, A.S. \etal 1999, 

\bibitem[Fruchter \etal (2006)]{f06} Fruchter, A.S. \etal 2006, Nature, 441, 463

\bibitem[Fynbo \etal (2003)]{f03} Fynbo, J.P.U., \etal 2003, A\&A, 406, L63

\bibitem[Hjorth \etal (2003)]{h03} Hjorth, J., \etal 2003, Nature, 423, 847

\bibitem[Jakobsson \etal (2006)]{j06} Jakobsson, P., \etal 2006, A\&A, 443, 897

\bibitem[Kobulnicky \& Kewley (2004)]{kk04} Kobulnicky, H.A., \& Kewley, L.J. 2004, ApJ, 617, 240

\bibitem[Lamb \& Reichart (2000)]{lr00} Lamb, D.Q., \& Reichart, D.E. 2000, ApJ, 536, 1

\bibitem[Le \& Dermer(2007)]{2007ApJ...661..394L} Le, T., \& Dermer, C.~D.\ 2007, \apj, 661, 394 

\bibitem[Le Floch \etal (2003)]{le03} Le Floch, E., 2003, A\&A, 400, 499

\bibitem[Li (2003)]{l07} Li, L.-X. 2007, arXiv:0710.3587

\bibitem[MacFayden \& Woosley (1999)]{mw99} MacFayden, A.I., \& Woosley, S.E. 1999, ApJ, 524, 262

\bibitem[Nagamine \etal (2006)]{n06} Nagamine, K., Fukugita, M., Cen, R., \& Ostriker, J.P. 2001, ApJ, 588, 497

\bibitem[Nagamine \etal (2006)]{n06} Nagamine, K., Ostriker, J.P., Fukugita, M., \& Cen, R. 2006, ApJ, 633, 881

\bibitem[Nakamura \& Umemura (2002)]{n06}Nakamura, F., \& Umemura, M. 2002, ApJ, 569, 549

\bibitem[Porciani \& Madau (2001)]{pm01} Porciani, C., \& Madau, P. 2001, ApJ, 548, 522

\bibitem[Prochaska (2006)]{p06} Prochaska, J.X. 2006, ApJ, 650, 272

\bibitem[Savaglio, Glazebrook, \& Le Borgne (2006)]{sgl06} Savaglio, S., Glazebrook, K., \& Le Borgne, D. 2006, in AIP Conf. Proc. Vol 836, Gamma-Ray Bursts in the Swift Era, eds. S. Holt, N. Gehrels, J. Nousek, Am. Inst. Phys., Melville, NY, p540

\bibitem[Stanek \etal (2003)]{s03} Stanek, K.Z., \etal 2003, ApJ, 591, L17

\bibitem[Stanek \etal (2006)]{s06} Stanek, K.Z., \etal 2006, Acta Astron. 56, 333

\bibitem[Totani (1999)]{t99} Totani, T. 1999, ApJ, 511, 41

\bibitem[Trac \& Cen (2007)]{tc07} Trac, H., \& Cen, R. 2007, astro-ph/0612406, ApJ, in press

\bibitem[Wijers \etal (1998)]{w98} Wijers, R.A., \etal 1998, MNRAS, 294, L13

\bibitem[Woosley \& Heger (2006)]{wh06} Woosley, S.E., \& Heger, A. 2006, ApJ, 637, 914

\end{thebibliography}
\end{document}